\documentclass{amsart}
\usepackage{amssymb}
\newtheorem{thm}{Theorem}
\newtheorem{cor}{Corollary}
\newtheorem{lem}[thm]{Lemma}

\newtheorem{defn}{Definition}
\newtheorem{rem}{Remark}

\newcommand{\scalar}[2]{\langle#1,#2\rangle}
\newcommand{\set}[1]{\left\{#1\right\}}
\newcommand{\Real}{\mathbb{R}}

\newcommand{\Complex}{\mathbb{C}}
\newcommand{\To}{\longrightarrow}
\newcommand{\image}[1]{\textrm{Im}(#1)}

\newcommand{\pars}[1]{\left(#1\right)}
\newcommand{\brck}[2]{\{#1,#2\}}
\newcommand{\jami}[2]{\sum\limits_{#1}^{#2}}
\newcommand{\smooth}[1]{C^\infty(#1)}

\newcommand{\gamomd}{\,\Longrightarrow\,}
\newcommand{\extder}{\textrm{d}}
\newcommand{\fromto}[3]{#1=#2,\ldots,#3}
\newcommand{\liealg}[1]{\mathfrak{#1}}
\newcommand{\img}[1]{\textrm{Image}(#1)}
\newcommand{\state}[1]{|#1\rangle}

\newcommand{\contgr}[1]{\mathcal{C}(#1)}
\newcommand{\vecf}[1]{\mathcal{F}(#1)}
\newcommand{\paths}[1]{\mathcal{P}(#1)}
\newcommand{\reach}[1]{\mathcal{R}(#1)}
\newcommand{\reachl}[1]{\hat{\mathcal{R}}(#1)}
\newcommand{\lie}[1]{\mathcal{L}ie(#1)}
\newcommand{\glie}[1]{\mathcal{L}(#1)}
\newcommand{\ad}[1]{\textrm{ad}_{#1}}
\newcommand{\su}[1]{\liealg{su}(#1)}

\newcommand{\remdiag}[2]{\widetilde{#1}^#2}
\newcommand{\contr}[1]{\mathcal{C}_{#1}}
\newcommand{\compl}[1]{#1^\Complex}
\begin{document}
\title{Geometric Control Methods for Quantum Computations}%
\author{Zakaria Giunashvili}%
\address{Department of Theoretical Physics,
         Institute of Mathematics,
         Georgian Academy of Sciences,
         Tbilisi, Georgia}%
\email{zaqro@gtu.edu.ge}%
\thanks{This work was partially supported by Nishina Memorial Foundation and was carried out during the author's visit in
Yokohama City University (Japan) in the framework of Nishina Memorial Foundation's Postdoctoral
Fellowship Program.}%
\date{January 28, 2003}%
\begin{abstract}
The applications of geometric control theory methods on Lie groups and homogeneous spaces
to the theory of quantum computations are investigated. These methods are shown to be
very useful for the problem of constructing an universal set of gates for quantum
computations: the well-known result that the set of all one-bit gates together with
almost any one two-bit gate is universal is considered from the control theory viewpoint.
\end{abstract}
\maketitle
\section{Introduction}
There are three main directions in the theory of quantum computations:
\begin{itemize}
    \item Algebraic: the problem of constructing a universal set of gates
    \item Physical (practical): the problem of constructing a physical device
    implementing a universal set of gates
    \item Algorithmic: inventing algorithms for quantum devices to solve various computational problems
    effectively, provided that a universal set of gates is already constructed (at least
    theoretically).
\end{itemize}
The first of these directions is pure algebraic (perhaps linear algebraic), since a
universal set of elementary gates is such subset of the group of unitary transformations
of some finite-dimensional Hilbert space which generates the entire group. In the case of
quantum computer there is a correspondence between computational tasks and unitary
transformations, and the quantum computer program solving the problem is a composition of
some elementary gates which equals to the unitary transformation corresponding to the
given problem. Therefore, any conventional (as classical so quantum) computational
process is discrete while the evolution of any physical system (as classical so quantum)
is continuous and is described with the corresponding Hamiltonian or Schr\"{o}dinger
equation. Hence, it can be stated that during a conventional computational process, when
we want to represent a given unitary transformation as a composition of some elementary
gates, we have a mechanical system with Hamiltonian controlled by some external means
(inputs) and we are able to implement the elementary gates by these inputs. The
computational process is considered as discrete because we are interested not with the
continuous evolution of the system but with the transformations at some finite moments of
time corresponding to the elementary gates. Nowadays there are several works where the
''discrete`` approach to the computational process is replaced by ''continuous``, when
the aim is to construct a given transformation of a system not by means of discrete
inputs generating the preset gates, but by continuous steering with some external
parameter-dependent Hamiltonian (very interesting approach is developed in \cite{Kieu1},
\cite{Kieu2}, \cite{Kieu3}). The geometric control theory is a tool which unifies these
two approaches and the aim of this work is to demonstrate the methods that can be used to
build a computational process (program) by parameter-dependent Hamiltonian and to
investigate what can be achieved by these methods.
\section{Dynamical Polysystems and Control Group}
\begin{defn}[Control Group, see \cite{Lobry}]
Let $I$ be any nonempty set. Consider the set of all finite sequences of type
$\pars{(t_1,i_1)\cdots(t_m,i_m)}$ where $(t_k,i_k)\in\Real\times I,\ k=1,\ldots,m$.
Introduce the following reduction rules:
\begin{enumerate}
    \item
        any term of the type $(0,i)$ is removed from the sequence
    \item
        if $i_k=i_{k+1}\equiv i$ then the pair of consecutive terms
        $(t_k,i_k)(t_{k+1},i_{k+1})$ is replaced by the single term $(t_k+t_{k+1},i)$
\end{enumerate}
It is clear that after a finite number of steps from any sequence can be obtained an
irreducible sequence (which can be empty and we denote such one by $0$). We denote the
set of all irreducible sequences by $\contgr{I}$ and call the control group over the set
$I$.
\end{defn}
\begin{rem}
Another point of view is to consider the control group as a quotient set of the set of
all sequences $\set{\pars{(t_1,i_1)\cdots(t_m,i_m)}}$ under the following equivalence
relation: two sequences are equivalent if after reduction they give one and the same
irreducible sequence.
\end{rem}
The following operations on the set $\contgr{I}$ justify the usage of term \emph{group}.
For any two sequences $s_1$ and $s_2$ from $\contgr{I}$ let $(s_1s_2)$ be the sequence
obtained by their concatenation and let the \emph{product} $s_1\cdot s_2$ be the
irreducible sequence obtained by the reduction of $(s_1s_2$). The operation
$(s_1,s_2)\mapsto s_1\cdot s_2$ together with the neutral element $0$ (the empty
sequence) defines a group structure on $\contgr{I}$.
\begin{rem}
The group $\contgr{I}$ is commutative only when $I$ is a one-element set.
\end{rem}
The elements of $\contgr{I}$ are called \emph{controls}. For any control
$s=\pars{(t_1,a_1)\cdots(t_n,a_n)}$ its inverse element in the group $\contgr{I}$ is the
control $s^{-1}=\pars{(-t_n,a_n)\cdots(-t_1,a_1)}$. A control
$\pars{(t_1,a_1)\cdots(t_n,a_n)}$ is said to be \emph{positive} (\emph{negative}) if
$t_i>0\ (t_i<0),\ \forall\ i\in\set{1,\ldots,n}$. Let us denote the set of all positive
(negative) controls by $\mathcal{C}^+(I)$ ($\mathcal{C}^-(I)$). In general
$\contgr{I}\neq\mathcal{C}^+(I)\cup\mathcal{C}^-(I)$ because there are many ``mixed''
type controls in $\contgr{I}$. It is clear that $\mathcal{C}^+(I)$ and $\mathcal{C}^-(I)$
are semigroups.

For any $\lambda\in\Real$ and $s=\pars{(t_1,a_1)\cdots(t_n,a_n)}\in\contgr{I}$ define a
control $\lambda\cdot s$ as $\lambda\cdot s=\pars{(\lambda t_1,a_1)\cdots(\lambda t_n,a_n)}$. %
Any subset $\set{s_1,\ldots,s_p}\subset\contgr{I}$ defines a mapping
$$
\Phi_{s_1\cdots s_p}:\Real^p\To\contgr{I},\quad\Phi_{s_1\cdots s_p}(\lambda_1,\ldots,\lambda_p)=\lambda_1s_1\cdot\lambda_2s_2\cdots\lambda_ps_p %
$$
After this we can introduce a topology on $\contgr{I}$ as the strongest topology for
which all the maps of type $\Phi_{s_1\cdots s_p}$ are continuous. Analogically can be
introduced a differential structure on $\contgr{I}$: a mapping $f:\contgr{I}\To\Real$ is
said to be smooth if the mapping $f\circ\Phi_{s_1\cdots s_p}:\Real^p\To\Real$ is smooth
for any $\Phi_{s_1\cdots s_p}$. The group operations in $\contgr{I}$ are continuous and
smooth with respect to the above defined topology and smooth structure.
\begin{defn}
For a smooth manifold $M$ a dynamical polysystem on $M$ controlled by $\contgr{I}$ is a
smooth action of the group $\contgr{I}$ on $M$:
$$
\contgr{I}\times M\ni(s,m)\mapsto sm\in M
$$
\end{defn}
\begin{defn}
The set $\contgr{I}x=\set{sx\ |\ s\in\contgr{I}}$ is called the \textbf{orbit} of the point $x\in M$. %
The set $\mathcal{C}^+(I)x=\set{sx\ |\ s\in\mathcal{C}^+(I)}$ is called the
\textbf{positive orbit} of $x$ and $\mathcal{C}^-(I)x=\set{sx\ |\ s\in\mathcal{C}^-(I)}$
-- the \textbf{negative orbit} of $x$.
\end{defn}
In this case any fixed element $a\in I$ gives a one-parameter subgroup $\mathcal{C}_a=\set{(t,a)\ |\ t\in\Real}\subset\contgr{I}$ %
which induces a one-parameter group of diffeomorphisms of $M$:
$$
\varphi^a_t:M\To M,\quad t\in\Real,\quad\varphi^a_t(x)=(t,a)x,\quad x\in M
$$
The latter itself corresponds to some smooth vector field $X^a$ on $M$. The family of
vector fields $\set{X^a\ |\ a\in I}$ is called the \emph{infinitesimal} transformations
of the dynamical polysystem. Conversely, if we have a family of vector fields $F$ on $M$
indexed by the elements of some set $I$, then we obviously have a dynamical polysystem on
$M$ controlled by $\contgr{I}$. In such cases we denote the control group together with
the corresponding dynamical polysystem by $\contgr{F}$. Thus, we can state that there is
a one-to-one correspondence between the dynamical polysystems on $M$ and the families of
smooth vector fields on $M$.
\section{The Structure of an Orbit of Dynamical Polysystem}
Let $F$ be a family of vector fields on the manifold $M$.
\begin{defn}
The closure of the family of vector fields $F$ is the family of vector fields defined as
$$
\Delta(F)=\set{sX\ |\ s\in\contgr{F},\ X\in F}
$$
In other words it is the minimal set of vector fields containing $F$ and invariant under
the action of the control group $\contgr{F}$.
\end{defn}
The \emph{dimension} of the dynamical polysystem $\contgr{F}$ at a point $x\in M$ is the
dimension of the subspace of the tangent space $T_x(M)$ generated by the family of vector
fields $\Delta(F)$. The main result about the structure of an orbit of dynamical
polysystem is based upon the following
\begin{lem}[see \cite{Sussmann1}]
For any point $x\in M$ the dimension of the dynamical polysystem $\contgr{F}$ is constant
through the orbit $\contgr{F}x$. If $m$ is the dimension of the dynamical polysystem
$\contgr{F}$ at the point $x$ then there exists a set of controls
$\set{s_1,\ldots,s_m}\subset\contgr{F}$ and a set of vector fields
$\set{X_1,\ldots,X_m}\subset F$ such that the mapping
$$
\phi^x:\Real^m\To\contgr{F}x,\quad\phi^x(t_1,\ldots,t_m)=(s_m\varphi^m_{t_m}s_m^{-1}\cdots s_1\varphi^1_{t_1}s_1^{-1}) %
$$
where $\set{\varphi^i_t\ |\ t\in\Real},\ i\in\set{1,\ldots,m}$ is the one-parameter group
of diffeomorphisms corresponding to $X_i$, is a local diffeomorphism at the point
$(0,\ldots,0)\in\Real^m$.
\end{lem}
Let $\exp(F)$ denote the group of diffeomorphisms of $M$ generated by the flows of the
elements of $F$:
$$
\exp(F)=\set{\exp(t_1X_1)\cdots\exp(t_nX_n)\ |\ t_i\in\Real,\ X_i\in F,\ n\in\mathbb{N}}
$$
and $\glie{F}$ be the minimal Lie subalgebra of $\vecf{M}$ containing $F$ and invariant
under the action of the group $\exp(F)$.

From the above lemma follows the following
\begin{thm}[see \cite{Sussmann1}]\label{orbit_theorem}
For any point $x$ in the orbit $\contgr{F}x_0,\ x_0\in M$, the mapping
$\phi^x:\Real^m\To\contgr{F}x_0$ gives a local coordinate system on $\contgr{F}x_0$. The
topology and the differential structure induced on the orbit by these coordinate system
coincide with the topology and the differential structure induced by the action of the
group $\contgr{F}$.
\end{thm}
and
\begin{thm}[Orbit Theorem, Nagano-Sussmann]\label{orbit_theorem_nagano-sussmann}For any point $x_0\in M$ the orbit
$\contgr{F}x_0\equiv\mathcal{O}_{x_0}$ is a connected immersed submanifold of $M$ and
$T_{x_0}(\mathcal{O}_{x_0})=\glie{F}_{x_0}$.
\end{thm}
\section{Control System and Accessibility Problem}\label{general_contrsys_section}
For a smooth manifold $M$ let $\vecf{M}$ be the space of smooth vector fields on $M$. We
shall generalize the notion of control group in the following way. Let $U$ be some
nonempty set (usually assumed to be a finite-dimensional smooth manifold). We denote by
$\paths{U}$ the set of all paths in $U$. At this point we do not specify what type of
paths we consider
$$
\paths{U}=\set{u:[0,T]\To U\ |\ T\geq0}
$$
The set $\paths{U}$ is a semigroup under the following standard operation: for
$u_1:[0,T_1]\To U$ and $u_2:[0,T_2]\To U$ let $u_2u_1:[0,T_1+T_2]\To U$ be
$$
(u_2u_1)(t)=
\begin{cases}
    u_1(t), & t\in[0,T_1]; \\
    u_2(t), & t\in[T_1,T_1+T_2]. \\
\end{cases}%
$$
A \emph{control system} on the manifold $M$ is the following data
\begin{itemize}
    \item a smooth map $X:U\To\vecf{M}$;
    \item a subspace $\contgr{U}$ of $\paths{U}$ (at this stage we do not specify the
    structure of $\contgr{U}$, but it is assumed to be a \emph{subsemigroup} of $\paths{U}$).
\end{itemize}
The elements of $\contgr{U}$ are called \emph{controls} and any $u\in\contgr{U},\ u:[0,T]\To U$ %
defines a dynamical system
\begin{equation}\label{control_equation}
\dot{m}=X^u(t)_{m(t)},\quad m(t)\in M,\quad t\in[0,T]
\end{equation}
where $X^u$ denotes the time-dependent vector field $X^u(t)=(X\circ u)(t)$ and
$X^u(t)_{m(t)}$ is the value of this vector field at the point $m(t)\in M$.
\begin{rem}
The solution of \ref{control_equation} is defined as the absolutely continuous map
$m:[0,T]\To M$ such that the equation is satisfied for almost every $t\in[0,T]$.
\end{rem}
For any given control $u:[0,T]\To U$ and an initial value $m(0)\equiv m_0$, we denote by
$\gamma(m_0,u,t),\ t\in[0,T]$ the solution (if exists) of \ref{control_equation}. If for
some $t\in[0,T]$ and $x\in M$ we have that $\gamma(m_0,u,t)=x$, then we state that the
control $u$ \emph{steers} the state $m_0$ into the state $x$ in time $t$, and the state
$x$ is \emph{reachable} from $m_0$ in time $t$. We denote by $\reach{x_0,t}$ the set of
points of $M$ reachable from the point $x_0\in M$ in time $t\in\Real$. Traditionally also
the following two sets are the subjects of investigation
$$
\reachl{x_0,t}=\bigcup\limits_{0\leq\tau\leq t}\reach{x_0,\tau}\textrm{ -- the set of
points reachable from }x_0\textrm{ in time }\leq t
$$
and
$$
\reachl{x_0}=\bigcup\limits_{0\leq\tau\leq\infty}\reach{x_0,\tau}\textrm{ -- the set of
points reachable from }x_0
$$
The aim of geometric control theory is the investigation of the structures of the sets
$\reach{x_0,t}$, $\reachl{x_0,t}$ and $\reachl{x_0}$. These problems are considered for
various types of the space of control parameters $U$ as well as the map $X:U\To\vecf{M}$
and the class of controls $\contgr{U}$. We assume that $U=\Real^m$ for some integer $m>0$
and consider the following three types of controls
\begin{itemize}
    \item $\mathcal{C}_u$ -- \emph{unrestricted controls}, which consists of locally bounded and
    measurable mappings from $[0,T]$ to $\Real^m,\ T\geq0$;
    \item $\mathcal{C}_r$ -- \emph{restricted controls}: the subset of $\mathcal{C}_u$
    consisting of the mappings with values in the cube $\set{(x_1,\ldots,x_m)\in\Real^m\ | \|x_i|\leq1,\ i=1,\ldots,m}$; %
    \item $\mathcal{C}_b$ -- \emph{``bang-bang''} controls, consisting of the all
    piecewise constant maps from $[0,T]$ to $\Real^m$.
\end{itemize}
It is clear that in the case of bang-bang controls the reachable set is independent of
the structure of the map $X:U\To\vecf{M}$ and depends only on the image of this map. The
set reachable from a point $m\in M$ in time $t\geq0$ for piecewise constant controls is
$$
\begin{array}{c}
\reach{m,t}=\{\exp(t_1X_1)\cdots\exp(t_kX_k)m\ |\\
\jami{i=1}{k}t_i=t,\ t_i\geq0,\ X_i\in\textrm{Image}(X),\ i=1,\ldots,k,\ k\in\mathbb{N}\}
\end{array}
$$
where $\exp(t\xi)$ denotes an element of the one-parameter group of diffeomorphisms
associated with a vector field $\xi$. In the case when the set $\textrm{Image}(X)$ is
symmetric in the sense that $\xi\in\textrm{Image}(X)\gamomd-\xi\in\textrm{Image}(X)$, we
have that the set reachable from any point $m\in M$ coincides with the orbit of the
dynamical polysystem on $M$ defined by the family of vector fields $\textrm{Image}(X)$.
So we can apply the classical Orbit Theorem (see Theorem \ref{orbit_theorem}) for the
structure of the set $\reachl{m},\ m\in M$.

Let $\lie{S}$ be the minimal Lie subalgebra of $\vecf{M}$ containing the subset
$S\subset\vecf{M}$. It is clear that for any point $x\in M$ the space $\lie{S}_x\subset T_x(M)$ %
is a subspace of the tangent space to the orbit of $x$. In some cases it coincides with
the tangent space of the orbit. Particularly, the following theorem is one of the most
important corollaries of the Orbit Theorem
\begin{thm}[Rashevsky-Chow]\label{Rashevski-Chow_theorem}
If $M$ is a connected smooth manifold and $\lie{S}_x=T_x(M)$ for every point $x\in M$,
then the orbit of the point $x$ under the action of the dynamical polysystem defined by
$S$ coincides with the entire manifold $M$.
\end{thm}
A set $S\subset\vecf{M}$ which satisfies the condition $\lie{S}_x=T_x(M), \forall\ x\in M$ %
is said to be \emph{completely nonholonomic} or \emph{bracket-generating}. As it was
mentioned if the set $S$ is symmetric then the orbits of the corresponding dynamical
polysystem and the reachable sets coincide. Hence, we can conclude that for symmetric
($S=-S$) nonholonomic family of vector fields the corresponding control system is
completely controllable.
\section{Two Notions of Controllability in the Case of Classical Hamiltonian System}
Sometimes it is useful to distinguish the following two notions of controllability
(reachability) as for classical so for quantum mechanical systems.
\begin{defn}[Transformation-Controllability]
We call a system \textbf{transformation-controllable} if every transformation of the
system can be obtained by applying of some finite set of appropriate controls in a finite
interval of time.
\end{defn}
\begin{defn}[State-Controllability]
We call a system \textbf{state-controllable} if for every pair of states of the system --
$s_1$ and $s_2$, there exists a finite set of controls that steers the system from the
state $s_1$ into the state $s_2$ in a finite time.
\end{defn}
In the case of classical system, if we assume that the set of all transformations is the
entire group of diffeomorphisms of a smooth manifold $M$ -- $Diff(M)$, and the control
system is the dynamical polysystem defined by some family of vector fields $F$, then the
transformation-controllability means that $\exp(F)=Diff(M)$; and the
state-controllability means that for any point $m\in M$: $\mathcal{O}_m=M$. These two
notions are not equivalent neither in classical nor in quantum cases. To demonstrate this
let us consider the following situation. Let $M$ be a symplectic manifold and the
dynamical polysystem $F$ be the family of all Hamiltonian vector fields on $M$:
$F=ham(M)$. Let $\brck{\ }{\ }$ denote the Poisson bracket corresponding to the
symplectic structure on $M$. According to the Darboux theorem for any point $m\in M$
there exists a local coordinate system $\set{p_i,\ q_i\ |\ i=1,\ldots,n}$ such that
$\brck{p_i}{q_j}=\delta_{ij}$. This implies that locally we have the set of Hamiltonian
vector fields
$\set{\frac{\partial}{\partial q_i},\ \frac{\partial}{\partial p_i}\ |\ i=1,\ldots,n}$ %
which forms a basis for $T_m(M)$. So we have that for any point $m\in M$ the set of
Hamiltonian vector fields forms a basis of $T_m(M)$. Therefore, by the theorem of
Rashevsky-Chow the family of all Hamiltonian vector fields is completely nonholonomic and
therefore the corresponding system is state-controllable. But the family $ham(M)$ is not
transfomation-controllable because the corresponding family of 1-parameter flows is
contained in the group of symplectic diffeomorphisms of the manifold $M$, and it is clear
that this group is just a part of the entire group of diffeomorphisms of $M$. The
classical problem of state-controllability for the family of all Hamiltonian vector
fields becomes more complicated if we extend the notion of state and consider not only
pure states (i.e. points), but also \emph{mixed states}. Let us recall the formal
definition of this notion which is rather well-known from statistical physics. The
realization of the algebra of observables as the algebra $\smooth{M}$ dictates the
realization of the space of states as the space of probability distributions, i.e., the
linear positive functionals $\rho:\smooth{M}\To\Real$ such that
$\phi>0\gamomd\rho(\phi)>0$ and $\int\limits_M\rho(p,q)\extder p\extder q=1$. Let us
denote the space of such functionals by $Stat(M)$. In other words, a state $\rho\in Stat(M)$ %
is a \emph{positive distribution} (generalized function) such that $\rho(1)=1$. The pure
state corresponding to a point $x\in M$ is the Dirac functional $\delta_x$:
$\delta_x(\phi)=\phi(x)$. Sometimes we shall use the scalar product notation
$\scalar{\rho}{\phi}$ for $\rho(\phi)$. Any smooth mapping of pure states $f:M\To M$ can
be extended to $f:Stat(M)\To Stat(M)$ as
$\scalar{f(\rho)}{\phi}=\scalar{\rho}{f^*(\phi)}$ where $f^*:\smooth{M}\To\smooth{M}$
denotes the mapping dual to $f$.
\begin{rem}
It is clear that the subspace $Stat(M)$ in the space of all generalized functions is
invariant under such mappings, i.e., every positive distribution is mapped to a positive
one and the property $\rho(1)=1$ is preserved.
\end{rem}
After this we extend the problems of state-controllability and reachability to the space
$Stat(M)$. First of all let us notice that the pure-state-controllabilty does not imply
the state-controllability in the extended state space. The simplest example is the
following: let $M$  be a simplectic manifold and $\omega$ be the symplectic form on it,
let $\alpha=a\cdot\omega^n,\ 2n=\dim(M)$ be the volume form normalized so that
$\int\limits_M\alpha=1$; take the state $\mathcal{E}$ which is a generalized function
defined as $\scalar{\mathcal{E}}{\phi}=\int\limits_M\phi\cdot\alpha,\ \phi\in\smooth{M}$.
As it was mentioned, for the dynamical polysystem generated by the family of all
Hamiltonian vector fields -- $Ham(M)$, the space of pure states is controllable but it is
not true for the space $Stat(M)$ because the state $\mathcal{E}$ is invariant under the
action of Hamiltonian flows. The reasonable classification of the entire space $Stat(M)$
from the point of view of reachability under the dynamical polysystem defined by $Ham(M)$
seems very difficult. As an example of application of Orbit Theorem we consider a
simplified version of this problem. Particularly, we consider the reachability problem
for the space of \emph{discrete} mixed states.

The space $Stat(M)$ is a convex space:
$$
\rho_1,\rho_2\in Stat(M)\gamomd c\rho_1+(1-c)\rho_2\in Stat(M),\ \forall\ c\in[0,1]
$$
Therefore, for any finite set of points $P\subset M$ and a function $s:P\To\Real$ such
that $\varphi>0$ and $\sum\limits_{x\in P}s(x)=1$, we can construct the distribution
$\rho(P,s)=\sum\limits_{x\in P}s(x)\delta_x\in Stat(M)$. Let us denote the subspace of
all such states in $Stat(M)$ by $Stat_0(M)$. From the property
$f(\delta_x)=\delta_{f(x)}$ follows that $Stat_0(M)$ is an invariant subset of $Stat(M)$.
Now we consider the reachability problem for the dynamical polysystem defined by $Ham(M)$
just for the space $Stat_0(M)$. Particularly we describe the quotient space
$Stat_0(M)/\sim$, where two states $\rho_1$ and $\rho_2$ from $Stat_0(M)$ are equivalent
iff there exists a finite set of Hamiltonians $\set{H_1,\ldots,H_k}\subset\smooth{M}$ and
a set of time intervals $\set{t_1,\ldots,t_k}\subset\Real$ such that
$\exp(t_1V_{H_1})\cdots\exp(t_kV_{H_k})\rho_1=\rho_2$, where $V_{H_i}$ denotes the
Hamiltonian vector field corresponding to the function $H_i$.

For any $\rho=\rho(P,s)\in Stat(M)$ we call the set of points $P\subset M$ the
\emph{support} of the distribution $\rho$ and denote it by $Supp(\rho)$. For any
$c\in\Real$ we call the number of elements in the set $s^{-1}(c)$ the \emph{multiplicity}
of $c$ and denote this number by $m(c)$. For a given $\rho(P,s)\in Stat(M)$, the set of
all pairs $\set{(c,m(c))\ |\ c\in\textrm{Image}(s)}$ we call the \emph{spectrum} of the
distribution $\rho$ and denote this set by $Spec(\rho)$. It is easy to verify that for
any diffeomorphism (or even bijection) $f:M\To M$ we have that
$f(\rho(P,s))=\rho(f(P),s\circ f)$. The latter implies that if two states
$\rho_1=\rho(P_1,s_1)$ and $\rho_2=\rho(P_2,s_2)$ are equivalent then
$Spec(\rho_1)=Spec(\rho_2)$. It turns out that the equality of spectra is also a
sufficient condition for the equivalency of two states in $Stat_0(M)$. Before we start to
prove the sufficiency let us proof the following auxiliary
\begin{lem}\label{lemma1}
Let $X_1,\ldots,X_k$ be tangent vectors at $k$ different points
$\set{m_1,\ldots,m_k}\subset M$. There exists such Hamiltonian $H\in\smooth{M}$ that for
the corresponding Hamiltonian vector field $V_H$ we have: $V_H(m_i)=X_i,\ i=1,\ldots,k$.
\end{lem}
\begin{proof}
For each point $m_i$ take an open neighborhood $U_i$ with canonical coordinate system
$\set{p^i_j,q^i_j\ |\ j=1,\ldots,k}$ so that $U_i\cap U_l=\emptyset$ when $i\neq l$.
Since the system of all Hamiltonian vector fields is completely nonholonomic, for each
$X_i$ there exists a function $f_i\in\smooth{M}$ such that $V_{f_i}(m_i)=X_i$ and
$f_i(M\setminus U_i)=0$. It is clear that the function $f_i$ does not ``affect'' the
function $f_l$ inside the neighborhood $U_l$ when $i\neq l$. Therefore the function
$H=\jami{i=1}{k}f_i$ is the desired Hamiltonian.
\end{proof}
\begin{thm}
Two states $\rho_1$ and $\rho_2$ from $Stat_0(M)$ are equivalent if and only if
$Spec(\rho_1)=Spec(\rho_2)$.
\end{thm}
\begin{proof}
Assume that the spectra of $\rho_1=\rho(P_1,s_1)$ and $\rho_2=\rho(P_2,s_2)$ are equal.
For $\rho_1$ consider the ordered set $(x_1,\ldots,x_n)$ where $n$ is the number of
elements in $P_1$, $x_i\in P_1,\ i=1,\ldots,n$, and $s_1(x_i)\leq s_1(x_{i+1})$. In the
same way we construct the ordered set $(y_1,\ldots,y_n)$ for $\rho_2$. From
$Spec(\rho_1)=Spec(\rho_2)$ follows that $s_1(x_i)=s_2(y_i),\ i=1,\ldots,n$. Thus the
problem is reduced to the existence of such element $g$ of a Hamiltonian flow that
$g(x_i)=g(y_i),\ \fromto{i}{1}{n}$.

For $n\in\mathbb{N}$ let $\remdiag{M}{n}$ be the subset of
$M^n=\underbrace{M\times\cdots\times M}_{n-times}$ defined as
$$
\remdiag{M}{n}=\set{(p_1,\ldots,p_n)\ |\ p_i\neq p_j\textrm{ when }i\neq j,\ i,j=1,\ldots,n}%
$$
$\remdiag{M}{n}$ is an open subset of $M^n$ and therefore it is a submanifold. The set
$M^n\setminus\remdiag{M}{n}$ is a union of submanifolds $M^n_{i,j}=\set{(p_1,\ldots,p_n)\ |\ p_i=p_j},\ i\neq j$. %
It is clear that $\dim(M^n_{i,j})=(n-1)\cdot\dim(M)$ and hence if $\dim(M)\geq 2$ and $M$
is connected then $\remdiag{M}{n}$ is also connected. But recall that $M$ is assumed to
be a symplectic manifold, so $\dim(M)\geq2$. Thus we have a connected manifold
$\remdiag{M}{n}$ and the control group generated by the Hamiltonian vector fields on $M$
acts on $\remdiag{M}{n}$ as
$$
g^m:\remdiag{M}{n}\To\remdiag{M}{n},\ g^m(x_1,\ldots,x_n)=(g(x_1),\ldots,g(x_n))
$$
where $g:M\To M$ is an element of some Hamiltonian flow. The question is: is this action
transitive or not? For a given Hamiltonian $H\in\smooth{M}$ the action of the
corresponding flow on $\remdiag{M}{n}$ is
$(x_1,\ldots,x_n)\mapsto(\exp(tH)x_1\ldots,\exp(tH)x_n),\ t\in\Real$; therefore, the
corresponding vector field on $\remdiag{M}{n}$ is $V_H^n=(V_H,\ldots,V_H)$. That is, for
any point $(x_1,\ldots,x_n)\in\remdiag{M}{n}$ the value of this vector field at this
point is $(V_H(x_1),\ldots,V_H(x_n))$. According to the Lemma \ref{lemma1} for any
tangent vector $(X_1,\ldots,X_n)$ at the point $(x_1,\ldots,x_n)$ there is a Hamiltonian
$H$ such that $V_H(x_1),\ldots,V_H(x_n)=(X_1,\ldots,X_n)$. This implies that the
dynamical polysystem on $\remdiag{M}{n}$ generated by the diagonal actions of the
Hamiltonian flows on $M$ is completely nonholonomic. Hence, any two states from
$Stat_0(M)$ with equal spectra are equivalent.
\end{proof}
Extending this to the entire set of states $Stat(M)$ just by direct analogy, we can see
that if two distributions $\omega_1$ and $\omega_2$ are equivalent then there exists such
diffeomorphism $g:M\To M$ that $g(Supp(\omega_1))=g(Supp(\omega_2))$. The problem of
existence of such diffeomorphism that maps one given closed set to another given closed
set is itself quite complicated.
\section{Control Systems on Lie Groups: Homogeneous and Affine Cases}
In the case of control systems on manifolds the main difficulty for controllability and
accessibility problems comes from the fact that even for a small number of control
parameters the control group is a subgroup of an infinite-dimensional Lie group. For
control systems on finite-dimensional Lie groups when we restrict ourselves with
right-invariant vector fields the problem becomes easier sometimes and can be solved
completely. In this section we review the cases of \emph{homogeneous} and \emph{affine}
controls which are discussed in the classical paper of Jurdjevich and Sussmann (see
\cite{Jurdjevich-Sussmann}). We start from the definitions of basic notions.

Let $G$ be a finite-dimensional Lie group and $\liealg{g}$ be its Lie algebra. According
to the general definition (see Section \ref{general_contrsys_section}) we need a set of
\emph{control parameters} $U$ which is assumed to be $\Real^m$ for some positive integer
$m$; a set of \emph{controls} $\mathcal{C}(U)$ which is a subset of the set of mappings
$u:\Real^+\To\Real^m$ and usually is the set of continuous, piecewise-continuous,
piecewise-constant or smooth maps; and a continuous map $H:\Real^m\To\vecf{G}$. In this
case we consider such $H$ that takes values in the subspace of right-invariant vector
fields on $G$. Thus, we can define $H$ as a continuous map $H:\Real^m\To\liealg{g}$ and
call it the \emph{Hamiltonian} (with hope that it will not cause any confusion). These
data defines a family of control-dependent differential equations on $G$
\begin{equation}\label{Lie_group_control_equation}
\dot{g}(t)=H(u(t))g(t),\quad g(t)\in G,\ g(0)=g_0,\ t\geq0
\end{equation}
where $u:\Real^+\To\Real^m$ runs through the space of controls $\mathcal{C}(\Real^m)$.
The right-hand side of the equation is a time-dependent right-invariant vector field on
$G$ defined by the Lie algebra elements $H(u(t)),\ t\in\Real^+$. As in general case the
solution of the equation \ref{Lie_group_control_equation} is absolutely continuous map
$g:\Real^+\To G$ that satisfies this equation for almost every $t\geq0$. For any given
control $u$ and initial state $g_0$ we denote by $\gamma(g_0,u,t),\ t\in\Real^+$ the
solution of the equation \ref{Lie_group_control_equation}.

If for some $\gamma(g_0,u,t)=g$ for some $u\in\mathcal{C}(\Real^m)$ and $t\geq0$ then we
say that the control $u$ \emph{steers} the state $g_0$ into the state $g$ in time $t$ and
the state $g$ is reachable from $g_0$ in time $t$. As in general case, we denote by:
\begin{itemize}
    \item $\reach{g_0,t}$ the set of all reachable in time $t$ elements $g\in G$;
    \item $\reachl{g_0,T}=\bigcup\limits_{0\leq t\leq T}\reach{g_0,t}$ -- the set reachable
    in time $\leq T$;
    \item $\reach{g_0}=\bigcup\limits_{0\leq t\leq\infty}\reach{g,t}$ -- the set
    reachable from $g_0$.
\end{itemize}
From the fact that the system is right-invariant easily follows the following properties
of the reachable sets
$$
\reach{g,t}=\reach{1,t}\cdot g,\quad\reachl{g,T}=\reachl{1,T}\cdot g,\quad\reach{g}=\reach{1}\cdot g %
$$
Therefore, for the right-invariant control systems on Lie groups it is sufficient to
study the set reachable from $1$.

In the above-mentioned work of Jurdjevich and Sussmann (\cite{Jurdjevich-Sussmann}) is
considered the case when the Hamiltonian $H:\Real^m\To\liealg{g}$ is an affine function
\begin{equation}\label{affine_Hamiltonian}
H(u_1,\ldots,u_m)=X_0+\jami{i=1}{m}u_i\cdot X_i,\quad\textrm{for fixed }\set{X_0,\ldots,X_m}\subset L%
\end{equation}
We make a brief review of the results and methods for such control systems, because it
turns out that they are useful not only for this specific class of systems but for other
classes of control systems.

Let us introduce the following three types of control parameters
\begin{itemize}
    \item $\contr{u}$ -- \emph{unrestricted controls}: locally bounded and measurable
    mappings from $\Real^+\To\Real^m$;
    \item $\contr{r}$ -- \emph{restricted controls}: a subset of $\contr{u}$ consisting
    of the mappings with values in the cube $\set{(x_1,\ldots,x_m)\in\Real^m\ |\ |x_i|\leq1,\ \fromto{i}{1}{m}}$; %
    \item $\contr{b}$ -- \emph{``bang-bang'' controls}: the subset of all piecewise
    constant mappings (in fact it suffices if $|x_i|=1,\ \fromto{i}{1}{m}$);
\end{itemize}
For these three classes of controls we use the notation $\contr{}$ without subscript when
some statement is formulated for each of them.

The following theorem from the classical theory of Lie groups is one of the important
tool for studying control systems on Lie groups.
\begin{thm}[Yamabe's Theorem, see \cite{Yamabe}]\label{Yamabe_theorem}
Let $G$ be a Lie group and let $H$ be a path-connected subgroup of $G$. Then $H$ is a Lie
subgroup of $G$.
\end{thm}
The following theorem states that for affine Hamiltonian (see \ref{affine_Hamiltonian})
the solution of the controls system \ref{Lie_group_control_equation} exists and moreover
is complete.
\begin{thm}[see \cite{Jurdjevich-Sussmann}]\label{exists_and_complete}
For every control $u\in\contr{}$ and $g_0\in G$ there exists a unique solution $g(t),\ t\in\Real^+$ %
of \ref{Lie_group_control_equation} defined for every $t\in[0,\ \infty]$ such that
$g(0)=g_0$.
\end{thm}
Let $\hat{\liealg{g}}$ be the Lie subalgebra of $L$ generated by $\set{X_0,\ldots,X_m}$
and $\hat{G}$ be the corresponding connected Lie subgroup of $G$.
\begin{thm}[see \cite{Jurdjevich-Sussmann}]\label{reachable_from_unit}
The set $\reach{1}$ is a semi-group in $G$ and if it is a group then it coincides with
$\hat{G}$.
\end{thm}
\begin{proof}
Let $g_1=\gamma(1,u_1,t_1)$ and $g_2=\gamma(1,u_2,t_2)$ be two elements of $G$ reachable
from $1$ applying the controls $u_1$ and $u_2$. Define a new control $v$ as
$$
v(\tau)=
\begin{cases}
    u_1(\tau) & \textrm{when } \tau\in[0,\ t_1] \\
    u_2(\tau-t_1) & \textrm{when }\tau>t_1
\end{cases}
$$
Since the system is right-invariant we have that $\gamma(1,v,t_1+t_2)=g_2g_1$.

Now assume that $\reach{1}$ is a group. As the space of controls is path-connected, the
subset $\reach{1}$ is also path-connected and therefore, according to the Yamabe's
theorem (see \ref{Yamabe_theorem}) it is a Lie subgroup of $G$. Since
$\reach{1}\subset\hat{G}$, we have that its Lie algebra $V$ is a subalgebra of
$\hat{\liealg{g}}$. Conversely, consider any element
$v(a_1,\ldots,a_m)=X_0+\jami{i=1}{m}a_iX_i\in\hat{\liealg{g}},\ a_i=\pm1$. By definition
of $\reach{1}$ we have that $\exp(tv)\in\reach{1}$ for $t\geq0$, but since $\reach{1}$ is
a group we have that $\exp(tv)\in\reach{1}$ for $t\leq0$ too. The elements
$v(a_1,\ldots,a_m)$, where $(a_1,\ldots,a_m)\in\Real^m$, are generators of
$\hat{\liealg{g}}$ which implies that $\hat{\liealg{g}}\subset V$. Since the groups
$\reach{1}$ and $\hat{G}$ are connected we have that $\reach{1}=\hat{G}$.
\end{proof}
An affine control system $X=(X_0,X_1,\ldots,X_m)$ is said to be \emph{homogeneous} if
$X_0=0$. In other words the Hamiltonian depends on control parameters linearly:
$H(u_1,\ldots,u_m)=\jami{i=1}{m}u_iX_i$.
\begin{thm}[see \cite{Jurdjevich-Sussmann}]
For homogeneous right-invariant control system on a Lie group $G$ the reachable set
$\reach{1}$, for any of the three classes of controls, is the connected Lie subgroup of
$G$ corresponding to the Lie algebra generated by $X$. If we use the class of
unrestricted controls $\contr{u}$ then for each $T>0$ we have that
$\reach{1,T}=\reach{1}$.
\end{thm}
\begin{proof}
Since the set $\reach{1}$ is a semigroup, according to the previous theorem it is
sufficient to show that if $g\in\reach{1}$ then $g^{-1}\in\reach{1}$. Assume that
$g=\gamma(1,u,t)$. Consider a control $v$ defined as
$$
v(\tau)=
\begin{cases}
    -u(t-\tau) & \textrm{when }\tau\in[0,t] \\
    u(\tau) & \textrm{when }\tau>t
\end{cases}
$$
it is easy to verify that $\gamma(1,v,t)=g^{-1}$.

The second part of the theorem states that any reachable state can be reached in
arbitrary short time interval if we are able to use the class of unrestricted controls.
Assume that $g=\gamma(1,u,t)$. For any given time $t_1>0$ define a control $v$ as
$$
v(\tau)=\frac{t}{t_1}\cdot u\pars{\frac{\tau t}{t_1}},\quad\tau\in\Real^+
$$
It is easy to show that $\gamma(1,v,t_1)=g$ which implies that
$\reach{1,t}\subset\reach{1,t_1}$ for arbitrary $t_1>0$.
\end{proof}
Summarizing the essential properties of the homogeneous control systems we have the
following
\begin{itemize}
    \item The set reachable from $1$ is a subgroup of $G$;
    \item The set reachable from $1$ is one and the same for three classes of controls:
    every $g\in G$ that can be reached from $1$ by means of unrestricted control can also
    be reached by means of only ``bang-bang'' control (possibly later);
    \item If we use the unrestricted controls then every reachable state $g\in G$ can be
    reached in arbitrary short time interval.
\end{itemize}
For the affine case the most important facts are that if the set reachable from $1$ is
dense in the Lie group $\hat{G}$ corresponding to the Lie algebra generated by the set
$\set{X_0,\ldots,X_m}$ then $\reach{1}=\hat{G}$ and if the subgroup $\hat{G}$ is compact
then always $\reach{1}=\hat{G}$ and $\exists\ t>0:\ \reachl{1,t}=\reach{1}$ (see
\cite{Jurdjevich-Sussmann}).

A control system of the type \ref{Lie_group_control_equation} is said to be
\emph{controllable from} $g\in G$ if $\reach{g}=G$. It is said to be \emph{controllable}
if it is controllable from every $g\in G$.

For right-invariant control system on Lie group the controllability from $1$ is
equivalent to the controllability because as we know $\reach{g}=\reach{1}\cdot g$. As it
follows from the reachability criteria a necessary condition for the controllability is
that the group $G$ is connected and $\hat{\liealg{g}}=\liealg{g}$. If $G$ is compact or
the system is homogeneous then this condition is sufficient too.

Though the class of affine (or homogeneous) Hamiltonians is very restricted, actually the
same methods can be used for the reachability problem when the Hamiltonian is a
continuous mapping $H:\Real^m\To\liealg{g}$ and as it will be evident from the further
discussions, for a compact Lie group more important is not the Hamiltonian itself, but
the class of controls and the image set of the Hamiltonian.
\section{Control System on Compact Lie Group with Continuous Hamiltonian}
This section is mainly based on the results of the work \cite{Domenico1} the main idea of
which is that if we weaken the condition for the Hamiltonian to be affine, or
homogeneous, and consider any continuous Hamiltonian, for compact Lie groups many
essential results are still true. More formally, consider a right-invariant control
system
$$
\dot{g}=H(u)g,\quad g\in G,\ u\in\contr{},\ g(0)=1
$$
where $G$ is a compact Lie group, $\liealg{g}$ is its Lie algebra,
$H:\Real^m\To\liealg{g}$ is a \emph{continuous} map, $\contr{}$ is the set of controls
consisting of all piecewise-continuous maps from $\Real^+$ to $\Real^m$, and $H(u)g$
denotes the tangent vector at $g\in G$ obtained from $H(u)\in\liealg{g}$ by the right
action of $g$. Let $\hat{\liealg{g}}$ be the Lie subalgebra of $\liealg{g}$ generated by
$\img{H}\subset\liealg{g}$ and $\hat{G}$ be the corresponding connected Lie subgroup in
$G$. The latter is the maximal integral submanifold for the right-invariant differential
system defined by $\hat{\liealg{g}}$. Since the control system is right-invariant and
each trajectory of this system is tangent to $\hat{\liealg{g}}$ at the point $1\in G$, we
have that they are contained in $\hat{G}$. Therefore, the set reachable from $1$ is a
subset of $\hat{G}$. After this, without restriction of generality it can be assumed that
$\hat{\liealg{g}}=\liealg{g}$ and $\hat{G}=G$.

Using the same method as for the case of affine Hamiltonian (see Theorem
\ref{reachable_from_unit}), it is easy to show that the reachable set $\reach{1}$ is a
semigroup.

To proceed further we need the following
\begin{thm}[see \cite{Jurdjevich-Sussmann}]\label{generators_exponent}
Let $G$ be a connected Lie group and $X=\set{X_1,\ldots,X_m}$ be a set of generators of
its Lie algebra $\liealg{g}$. Then every $g\in G$ is a finite product of elements of the
form $\exp(tX_i),\ t\in\Real,\ X_i\in X$.
\end{thm}
\begin{proof}
The set of all finite products of the form $\exp(tX_i)$ is a path-connected subgroup $G'$
of $G$. According to the Yamabe's theorem (see \ref{Yamabe_theorem}) $G'$ is a Lie
subgroup of $G$. Since $G'$ contains the elements of the type $\exp(tX_i)$ its Lie
algebra $\liealg{g}'$ contains the set $X$, therefore it contains $\liealg{g}$.
Consequently, we have that $\liealg{g}'=\liealg{g}$. As the subgroups $G$ and $G'$ are
connected with one and the same Lie algebra, they coincide.
\end{proof}
Since $\liealg{g}$ is the Lie algebra generated by $\img{H}$, we can take the set of
generators $X$ from $\img{H}$. For the reachable set $\reach{1}$ the elements
$\exp(tX_i)$ are constructed by using of only positive time $t$, while for the elements
of the group $G$ can be used any $t\in\Real$.
\begin{lem}[see \cite{Domenico1}]\label{reachable_is_dense}
The set reachable from $1$ is dense subset of $G$.
\end{lem}
\begin{proof}
Consider an element $g\in G$ such that $g=\prod\limits_{i}\exp(t_iX_{p_i})$ where all
$t_i$-s are negative. We have that $h=g^{-1}$ is an element of $\reach{1}$ (because all
times become positive). Consider the sequence $h^n,\ n\in\mathbb{N}$. Since the group $G$
is compact this sequence contains a convergent subsequence $\set{h^{n_k}\ |\ k\in\mathbb{N}}$. %
We can assume that the sequence of subscripts $\set{n_k}$ is increasing. Take the
sequence $\set{h'_k=h^{n_{k+1}-n_k-1}}$. Since $n_k$ is assumed to be increasing, we have
that $n_{k+1}-n_k-1\geq0$ and therefore $\set{h'_k}\subset\reach{1}$. But obviously
$\lim\limits_{k\rightarrow\infty}h'_k=h^{-1}=g$. We obtain that the elements
$g=\prod\limits_{i}\exp(t_iX_{p_i})$ with only negative $t_i$-s can be approximated by
the elements of $\reach{1}$ which implies that $\reach{1}$ is dense in $G$.
\end{proof}
The following lemmas help us to go further and show that the set reachable from $1$
coincides with $G$. For $g\in G$ and $V\in\liealg{g}$ let $Ad(g)V$ be the element of
$\liealg{g}$ obtained by the adjoint action of $g$.
\begin{lem}[see \cite{Domenico2}]\label{lemma2}
Let $X=\set{X_1,\ldots,X_m}$ be a set of generators of the Lie algebra $\liealg{g}$. If
$m=\dim(\liealg{g})$ then for any $\epsilon>0$ there exist two elements $X_p$ and $X_q$
in
$X$ and a time $\tau,\ |\tau|<\epsilon$ such that the element $X_{m+1}=Ad(\exp(\tau X_p))X_q$ %
is linearly independent from $\set{X_1,\ldots,X_m}$.
\end{lem}
\begin{proof}
From $m<\dim(\liealg{g})$ follows that at least for one pair of elements $X_p$ and $X_q$,
their commutator $[X_p,X_q]$ is linearly independent from the set $X$. If we assume that
$Ad(\exp(tX_p))X_q=\jami{i=1}{m}a_i(t)X_i$, then the derivation of this equality by $t$
gives $[X_p,X_q]=\jami{i=1}{m}\dot{a}_i(0)X_i$ which is a contradiction.
\end{proof}
Applying the same procedure to the extended set of generators
$\set{X_1,\ldots,X_m,X_{m+1}}$ and etc., we can formulate the following
\begin{lem}\label{basis_from_generators}
The set of generators $\set{X_1,\ldots,X_m}$ can be extended to a basis of $\liealg{g}$
--
$\set{X_1,\ldots,X_m,X_{m+1},\ldots,X_{m+p}}$ so that each element $X_{m+i},\ \fromto{i}{1}{p}$ %
can be written as
$$
X_{m+i}=Ad(\exp(t_qX_{l_q})\cdots\exp(t_1X_{l_1}))X_{l_0}
$$
where $\set{X_{l_0},\ldots,X_{l_q}}$ is a subset of $\set{X_1,\ldots,X_m}$.
\end{lem}
Moreover, in the work of D. D'Alessandro (see \cite{Domenico2}) there is calculated some
upper bound for the number of factors in $\exp(t_qX_{l_q})\cdots\exp(t_1X_{l_1})$.
\begin{lem}
The set reachable from $1$ contains an open subset of $G$.
\end{lem}
\begin{proof}
Let $\set{X_1,\ldots,X_n}$ be a basis of $\liealg{g}$ as a vector space. Consider the
mapping $F:\Real^n\To G$
$$
F(t_1,\ldots,t_n)=\prod\limits_{i=1}^n\exp(t_iX_i),\quad(t_1,\ldots,t_n)\in\Real^n
$$
It follows from the inverse function theorem that $F$ is a local diffeomorphism on a
neighborhood of $0\in\Real^n$. This implies that for sufficiently small $\delta>0$ and
some $\epsilon\in(0,\delta)$, the open neighborhood of $(\delta,\ldots,\delta)\in\Real^m$
--
$$
\set{(t_1,\ldots,t_m)\ |\ \delta-\epsilon<t_i<\delta+\epsilon,\ i=1,\ldots,m}\subset\Real^m %
$$
is diffeomorphic to some open neighborhood $V\subset G$ of $g=F(\delta,\ldots,\delta)\in G$. %
According to Lemma \ref{basis_from_generators} there is a basis of $\liealg{g}$ ---
$\set{X_1,\ldots,X_n}$, where the first $m$ elements are generators from $\img{H}$ and
after the subscript $m$ the elements are of the form
$$
X_{m+k}=g_kX_{p_k}g_k^{-1},\quad p_k\in\set{1,\ldots,m},\ g_k\in G,\ \fromto{k}{1}{n-m}
$$
Regarding this, we obtain that the mapping $F$ is of the form
$$
\begin{array}{c}
F(t_1,\ldots,t_n)=e^{t_1X_1}\cdots e^{t_mX_m}\underbrace{g_1}e^{t_{m+1}X_{p_1}}\underbrace{g_1^{-1}g_2}e^{t_{m+2}X_{p_2}}g_2^{-1}\cdots\\ %
\cdots g_ke^{t_{m+k}X_{p_k}}\underbrace{g_k^{-1}g_{k+1}}\cdots g_{n-m}e^{t_nX_{p_{n-m}}}\underbrace{g_{n-m}^{-1}} %
\end{array}
$$
Consider the following terms of this product:
$$
g'_1=g_1,\ g'_2=g_1^{-1}g_2,\ldots,g'_k=g_{k-1}^{-1}g_k,\ldots,g'_{n-m}=g_{n-m}^{-1}
$$
They are not necessarily in $\reach{1}$, but since $\reach{1}$ is dense in $G$ we can
select elements of $\reach{1}$ -- $h_1,\ldots,h_{n-m}$ arbitrarily close to
$g'_1,\ldots,g'_{n-m}$
$$
h_1\approx g'_1,\ldots,h_{n-m}\approx g'_{n-m}
$$
and consider the mapping $\tilde{F}:\Real^n\To G$
$$
\tilde{F}(t_1,\ldots,t_n)=e^{t_1X_1}\cdots e^{t_mX_m}h_1e^{t_{m+1}X_{p_1}}h_2\cdots e^{t_nX_{p_{n-m}}}h_{n-m} %
$$
The latter is close to $F$ and therefore maps some open neighborhood of
$(\delta,\ldots,\delta)$ onto an open neighborhood of
$g'=\tilde{F}(\delta,\ldots,\delta)$. But for positive $t_i$-s we have that
$\tilde{F}(t_1,\ldots,t_n)$ is an element of $\reach{1}$. Hence, the set $\reach{1}$
contains an open subset of $G$.
\end{proof}
\begin{lem}
If the set reachable from $1$ contains some open subset of $G$ then it contains an open
neighborhood of $1$.
\end{lem}
\begin{proof}
Let $V$ be an open subset of $G$ contained in $\reach{1}$. Consider the set
$V^{-1}=\set{g^{-1}\ |\ g\in V}$. It is clear that $V^{-1}$ is also an open subset in
$G$, though it is not necessarily contained in $\reach{1}$. Since $\reach{1}$ is dense in
$G$ (see \ref{reachable_is_dense}), the set $V^{-1}$ contains at least one element
$h\in\reach{1}$. Consider the set $h\cdot V$. As $h\in V^{-1}$ clearly the set $h\cdot V$
contains $1$ and is open in $G$. Because $h\in\reach{1}$ and $V\subset\reach{1}$ and
$\reach{1}$ is a semigroup, we have that $h\cdot V\in\reach{1}$. Therefore, $\reach{1}$
contains an open neighborhood of $1$.
\end{proof}
Hence, we have that $\reach{1}$ is a semigroup in $G$ and contains an open neighborhood
of $1\in G$. These implies that $\reach{1}=G$.

To summarize, we can conclude that \emph{in the case of continuous Hamiltonian and
unrestricted controls on a compact Lie group, the set reachable from $1$ is the connected
Lie subgroup corresponding to the Lie algebra generated by $\img{H}$ (regardless of the
structure of the Hamiltonian $H:\Real^m\To\liealg{g}$)}.
\begin{rem}
Actually, during the proofs of the lemmas we use only ``bang-bang'' controls which is in
accordance with the fact that the reachable sets for the unrestricted and ``bang-bang''
controls are one and the same.
\end{rem}
\section{Minimal Set of Generators for $\liealg{su}(n)$}

As it was discussed in the previous section the controllability for right-invariant
``bang-bang'' (or piecewise continuous) systems on compact Lie groups is completely
described by the differential systems defined by the set of generators contained in the
image of the corresponding Hamiltonian. Hence, the question about a minimal set of
generators for a given Lie algebra (and the corresponding Lie group too) is natural. For
this question the following classical results of Kuranishi (see \cite{Kuranishi}) and
more resent results from \cite{Jurdjevic}, \cite{Altafini}, \cite{BreGe} and
\cite{Schirmer} are quite useful. In this section we give a review of some of them. First
let us recall the general structure of a semi-simple Lie group and the corresponding Lie
algebra.

Let $G$ be a compact Lie group and $\liealg{g}$ be its Lie algebra. The \emph{rank} of
$G$ is defined as the dimension of the maximal abelian subalgebra $\liealg{h}$ of
$\liealg{g}$. This number coincides with the dimension of the maximal torus (i.e., the
maximal connected abelian subgroup) $H$ in $G$.

All irreducible representations of $H$ are 1-dimensional and are described by the
elements of the group of \emph{characters} $\hat{H}=Hom(H,\Complex^*)$, where
$\Complex^*$ denotes the group $\Complex\setminus\set{0}$. Let $\compl{\liealg{g}}$
denote the complexification of $\liealg{g}$:
$\compl{\liealg{g}}=\liealg{g}\otimes\Complex$. Any representation $R$ of $G$ in a
complex vector space $V$ defines a representation $\dot{R}$ of $\compl{\liealg{g}}$ in
$V$. The restriction of $R$ on $H$ gives the decomposition $V=\bigoplus\limits_\lambda V_\lambda$, %
where $\lambda$-s are some elements of $(\compl{\liealg{h}})^*$ and %
$V_\lambda=\set{v\in V\ |\ R(\exp(h))v=\exp(\lambda(h))\cdot v,\ h\in\compl{\liealg{h}}}$. %

A vector $\lambda\in(\compl{\liealg{h}})^*$ is called a \emph{weight} of the
representation $R$ if $\dim(V_\lambda)>0$. The dimension of $V_\lambda$ is called the
\emph{multiplicity} of the weight $\lambda$.

The adjoint action of $G$ on $\compl{\liealg{g}}$ is a representation of $G$ on
$\compl{\liealg{g}}$ the restriction of which on $H$ gives the decomposition
$$
\compl{\liealg{g}}=\compl{\liealg{h}}+\bigoplus\limits_{\alpha\neq0}\liealg{g}_\alpha
$$
where $\alpha\in(\compl{\liealg{h}})^*$ and
$$
\liealg{g}_\alpha=\set{X\in\compl{\liealg{g}}\ |\ [Y,X]=\alpha(Y)\cdot X,\ \forall\ Y\in\compl{\liealg{h}}}%
$$
The nonzero weights of the adjoint representation are called the \emph{roots} of the
group $G$ with respect to the maximal torus $H$. Let $\Delta$ denotes the set of roots.

Let $\scalar{\ }{\ }$ be a $G$-invariant scalar product on $\liealg{g}$. For instance, it
can be the Killing form $\scalar{X}{Y}=tr(ad_X\circ ad_Y)$. This scalar product can be
linearly extended to $\compl{\liealg{g}}$. The algebra $\liealg{g}$ is said to be
\emph{semi-simple} if the scalar product $\scalar{\ }{\ }$ on $\liealg{g}$ is
non-degenerated. In this case the group $G$ is said to be \emph{semi-simple} too. Further
we assume that the group $G$ (and hence, the Lie algebra $\liealg{g}$) is semi-simple.

Notice that all roots are pure imaginary: $\Delta\subset i\liealg{h}^*$ and for any
$\alpha\in\Delta$ we have that $\bar{\liealg{g}}_\alpha=\liealg{g}_{-\alpha}$. The latter
implies that $\Delta=-\Delta$. Besides that $\dim(\liealg{g}_\alpha)=1$. The following
theorem summarizes the structure of a semi-simple Lie algebra.
\begin{thm}
Let $G$ be a connected compact Lie group with dimension $n$ and rank $l$. Then $G$ has
$2m$ roots $\set{\pm\alpha_k\ |\ \fromto{k}{1}{m}}$, where $n=l+2m$. The group $G$ is
semi-simple if and only if it has $l$ linearly independent roots. If $G$ is semi-simple
we can choose vectors $e_\alpha\in l_\alpha,\ \alpha\in\Delta$ so that
$$
[e_\alpha,e_{-\alpha}]=h_\alpha\in\liealg{h}
$$
and
$$
[e_\alpha,e_\beta]=
\begin{cases}
    0 & \textrm{ if } \alpha+\beta \textrm{ is not a root}\\
    c_{\alpha\beta}e_{\alpha+\beta} & \textrm{ if } \alpha+\beta \textrm{ is a root}
\end{cases}
$$
The set $\set{h_\alpha\ |\ \alpha\in\Delta}$ contains a basis for $\liealg{h}$ and the
vectors $e_\alpha$ are called the \textbf{root vectors}.
\end{thm}
An element $h\in\liealg{h}$ is said to be \emph{regular} if $\set{X\in\liealg{g}\ |\ [h,X]=0}=\liealg{h}$. %
Otherwise the element $h$ is said to be \emph{singular}. It is clear that
$h\in\liealg{h}$ is regular if and only if $\alpha(h)\neq0$ for each $\alpha\in\Delta$.

An element $h\in\liealg{h}$ is said to be \emph{strongly regular} if
$\alpha_1(h)\neq\alpha_2(h)$ for any two distinct roots $\alpha_1$ and $\alpha_2$. From
the property $\Delta=-\Delta$ follows that if $h$ is strongly regular then it is regular
too.

Based on these data about the structure of a semi-simple Lie algebra we present the proof
of the following classical theorem (see \cite{Kuranishi})
\begin{thm}\label{kuranishi_theorem}
Let $L$ be a semi-simple Lie algebra. There exist two elements in $\compl{L}$ which
generate the entire Lie algebra $\compl{L}$.
\end{thm}
\begin{proof}
Let $A$ be a maximal abelian subalgebra of $L$ and
$\compl{L}=\compl{A}+\bigoplus\limits_{\alpha\in\Delta}l_\alpha$ be the corresponding
root space decomposition of $L$. Let one element we are looking for be any strongly
regular element $h\in A$ and the other one be the vector
$e=\jami{\alpha\in\Delta}{}e_\alpha$. Consider the vectors
$$
\begin{array}{l}
    e_0=e \\
    e_1=[h,e_0]=\jami{\alpha\in\Delta}{}\alpha(h)e_\alpha \\
    \vdots \\
    e_{k+1}=[h,e_k]=\jami{\alpha\in\Delta}{}\alpha(h)^ke_\alpha \\
    \vdots \\
    e_{2m-1}=[h,e_{2m-2}]=\jami{\alpha\in\Delta}{}\alpha(h)^{2m-1}e_\alpha
\end{array}
$$
where $2m$ is the number of elements in $\Delta$. Since the numbers $\alpha(h),\
\alpha\in\Delta$ are different, the matrix
$$
\begin{pmatrix}
    1                   & \cdots    & 1                     \\
    \alpha_1(h)         & \cdots    & \alpha_{2m}(h)        \\
    \vdots              & \cdots    & \vdots                \\
    \alpha_1(h)^{2m-1}  & \cdots    & \alpha_{2m}(h)^{2m-1}
\end{pmatrix}
$$
where $\set{\alpha_1,\ldots,\alpha_{2m}}=\Delta$, is non-degenerate and therefore, the
set of vectors $\set{e_0,\ldots,e_{2m-1}}$ is a basis for
$\bigoplus\limits_{\alpha\in\Delta}l_\alpha$. Thus, from the vectors
$e_0,\ldots,e_{2m-1}$ we can construct the root vectors $e_\alpha,\ \alpha\in\Delta$ and
then use the vectors $[e_\alpha,e_{-\alpha}],\ \alpha\in\Delta$ to construct the basis
for $\compl{A}$.
\end{proof}
Notice that the above theorem concerns not the Lie algebra $L$ itself, but its
complexification. Now we consider the ways for constructing the pair of generators for a
Lie algebra itself. We concentrate on the Lie algebra $\liealg{su}(n)$. First we review
the structure of its complexification in the framework of root space decomposition.

Let $A$ be any element of $\su{n}$. We can take $A$ as a diagonal matrix, otherwise we
can diagonalize it. So, we have
$A=diag(\lambda_1,\ldots,\lambda_n),\ \jami{i=1}{n}\lambda_j=0,\ \lambda_j\in i\cdot\Real$. %
Let $H$ be the maximal abelian subalgebra of $\compl{\su{n}}$ containing $A$, and let
$H'$ be the zero eigenspace of $\ad{A}$:
$$
H'=\set{X\in\compl{\su{n}}\ |\ [A,X]=0}
$$
It is clear that
$$
H=H'\ \Longleftrightarrow \lambda_p\neq\lambda_q,\textrm{ when }p\neq q
$$
Therefore, the element $A$ is regular if and only if its eigenvalues $\lambda_j,\ \fromto{j}{1}{n}$ %
are distinct. We assume that the latter is true for $A$ and thus $H$ is the subgroup of
all diagonal (with respect to the basis in which $A$ is diagonal) traceless matrices. We
have $n$ linear functions $\lambda_j:H\To\Complex,\ \fromto{j}{1}{n}$, where for any
$X\in H$ we have $X=diag(\lambda_1(X),\ldots,\lambda_n(X))$. Let $\mathcal{E}_{pq}$ be
the matrix with $(p,q)$-entry equal to 1 and other entries equal to 0. It is clear that
for any $X\in H$ we have $[X,\mathcal{E}_{pq}]=\alpha_{pq}(X)\cdot\mathcal{E}_{pq}$,
where $\alpha_{pq}(X)=\lambda_p(X)-\lambda_q(X)$. Therefore, the matrices
$\mathcal{E}_{pq}\in\compl{\su{n}}$ can be considered as the root vectors and the
functionals $\alpha_{pq}=\lambda_p-\lambda_q$ as the roots. Each root vector has the
multiplicity 1 if and only if $\alpha_{pq}\neq\alpha_{ls}$ for $(p,q)\neq(l,s)$. It is so
when $H$ contains such matrix $B$ that
$\lambda_p(B)-\lambda_q(B)\neq\lambda_l(B)-\lambda_s(B)$ for $(p,q)\neq(l,s)$. Recall
that such element $B$ is said to be strongly regular. We assume that the matrix $A$
itself is strongly regular.

The roots $\alpha_{12},\alpha_{23},\ldots,\alpha_{n-1n}$ are called the \emph{fundamental
roots} since the others are obtained as sums of pairs of them. The basis for
$\compl{\su{n}}$ corresponding to the roots is
$$
\set{E_k=\mathcal{E}_{kk}-\mathcal{E}_{k+1k+1}\ |\ \fromto{k}{1,}{n-1}}\cup\set{\mathcal{E}_{pq}\ |\ \fromto{p,q}{1}{n},\ p\neq q} %
$$
This basis is known as the Wayl basis for $\compl{\su{n}}$. It is clear that the elements
of this basis are not in $\su{n}$ but such basis can be constructed from the Wayl basis
as
$$
\set{h_k=i\cdot E_k}\cup\set{U_{pq}=\mathcal{E}_{pq}-\mathcal{E}_{qp}}\cup\set{V_{pq}=i\cdot(\mathcal{E}_{pq}+\mathcal{E}_{qp})} %
$$
It is worth to notice the following commutating relations: for any diagonal $A=i\cdot
h\in\su{n}$ we have
\begin{equation}\label{comm_relations}
\begin{array}{l}
[A,U_{pq}]=\alpha_{pq}(A)V_{pq},\ [A,V_{pq}]=-\alpha_{pq}(A)U_{pq}\\
\\
\left[U_{pp+1},V_{pp+1}\right]=2h_p
\end{array}
\end{equation}
Our further discussion in this section is in accordance with \cite{Altafini}. Hence, we
have a matrix $A\in\su{n}$ which is a regular element (i.e., its eigenvalues have
multiplicity 1) and therefore, the associated Cartan subalgebra in $\su{n}$ is the set of
all traceless diagonal matrices. Let $B$ be another element of $\su{n}$. Consider the
graph $\Gamma_{A,B}$ defined as follows: $\Gamma_{A,B}$ has $n$ nodes
$\set{\state{1},\ldots,\state{n}}$ and the nodes $\state{p}$ and $\state{q}$ are joined
if and only if the $(p,q)$-entry of the matrix $B$ is nonzero. The graph is said to be
\emph{connected} if for all pairs of nodes there exists an oriented path connecting them.
The following two theorems are very important for the controllability of affine systems
on $\su{n}$ (see \cite{Altafini})
\begin{thm}
For $A,B\in\su{n}$ if $A$ is diagonal, a \textbf{necessary} condition that the pair
$(A,B)$ generates the entire $\su{n}$ is that the graph $\Gamma_{A,B}$ is connected.
\end{thm}
\begin{thm}
Given a pair of matrices $A,B\in\su{n}$, assume that the graph $\Gamma_{A,B}$ is
connected. If $A$ is \textbf{strongly regular} then the pair $(A,B)$ generates the entire
Lie algebra $\su{n}$.
\end{thm}
The first step of the proof (as in the case of the Theorem \ref{kuranishi_theorem}) is to
consider the set of matrices $A,\ [A,B],\ldots,[\underbrace{A,\ldots,[A}_{n^2-n-1},B]]$
and then use the commutating relations \ref{comm_relations} to construct the remaining
$n-1$ elements of the basis by the commutators between them.
\section{Quantum Gates: Recursive Construction of Generators}
Regardless of these, in quantum computation, it is important to construct the set of
generators (gates) recursively: assuming that we have the complete controllability of
some subsystem of a large system and the task is to construct an optimal extension of the
controllability to the entire system. Our further discussions makes more exact this
question.

Let $G$ be a compact Lie group with Lie algebra $\liealg{g}$; $\liealg{g}_0$ be a Lie
subalgebra of $\liealg{g}$ and $G_0$ be the corresponding connected Lie subgroup in $G$.
Assume that we have a set of generators of $\liealg{g}_0$, thus the set reachable from
$1\in G$ by using these generators is $G_0$. The problem is to find a minimal extension
of the given set of generators of $\liealg{g}_0$ to a set of generators of $\liealg{g}$.
This problem is actual for control systems used for quantum computations, because,
usually systems of universal quantum gates are constructed hierarchically, by adding new
gates to existing lower-dimensional subsystems.

Consider the adjoint action of the subgroup $G_0$ on the Lie algebra $\liealg{g}$. Since
$\liealg{g}_0$ is the Lie algebra of $G_0$, it is invariant under this action and thus we
have the following decomposition $\liealg{g}=\liealg{g}_0\oplus L_1\oplus\cdots\oplus L_p$ %
where each $L_i$ is invariant and irreducible for the adjoint action of $G_0$.
\begin{rem}
For instance, we can take the orthogonal complement of $\liealg{g}_0$ under a scalar
product invariant for the adjoint action. This space itself is invariant for the adjoint
action of $G_0$, and then take its decomposition on irreducible components.
\end{rem}
\begin{lem}\label{lemma_extension_of_generators}
Consider a set $A=\liealg{g}_0\cup\set{X_1,\ldots,X_p}$ where $X_i\in L_i\setminus\set{0},\ \fromto{i}{1}{p}$. %
The set $A$ generates the entire Lie algebra $\liealg{g}$.
\end{lem}
\begin{proof}
Let $\liealg{g}'$ be the Lie algebra generated by the set $A$. Since $A$ is a subset of
$\liealg{g}'$, the space $\liealg{g}'$ is invariant under the adjoint action of the
subgroup $G_0$. For each $\fromto{i}{1}{p}$ the intersection $\liealg{g}'\cap L_i$ is
nontrivial, because it contains at least $X_i$. Since the subspaces $\liealg{g}'$ and
$L_i$ are invariant for the adjoint action of $G_0$, their intersection is also
invariant. But by assumption the adjoint action of $G_0$ on each $L_i$ is irreducible,
which implies that $L_i$ is a subspace of $\liealg{g}'$ for each $\fromto{i}{1}{p}$ and
consequently $\liealg{g}'=\liealg{g}$.
\end{proof}
After this it is still not clear is the set $\set{X_1,\ldots,X_p}$ a minimal complement
of $\liealg{g}_0$ to a set of generators of $\liealg{g}$ or not. It is so when $p=1$, but
in general the answer is negative. To illustrate this, consider the
case when $\dim(L_i)=1,\ \fromto{i}{1}{p}$. Let us select nonzero vectors $e_i\in L_i,\ \fromto{i}{1}{p}$. %
The set $\set{e_1,\ldots,e_p}$ is a basis for $L=\bigoplus\limits_{i=1}^pL_i$, and we
have that $[x,e_i]=\lambda_i(x)\cdot e_i,\ x\in\liealg{g}_0$, where $\lambda_i$ is the
character of the representation of $G_0$ on $L_i$ via the adjoint action. Consider the
vector $e=\jami{i=1}{p}e_i$. If we are able to find such elements $x_1,\ldots,x_p$ in
$\liealg{g}_0$ that the matrix $\left(a_i^j=\lambda_i(x_j)\right)_{i,j=1}^p$ is
non-degenerate then the vectors $[x_1,e],\ldots,[x_p,e]$ form a basis for $L$ and
therefore we need just one element $e$ which together with $\liealg{g}_0$ gives a set of
generators of $\liealg{g}$, and of course it is a minimal complement for any set of
generators of $\liealg{g}_0$ to a set of generators of $\liealg{g}$.

The method described in the Lemma \ref{lemma_extension_of_generators} is in agreement
with the standard method of constructing the universal set of gates for quantum
computation. Now we describe a method of recursive construction of generators follow the
ideas discussed in \cite{Brylinski}. Let $X$ and $Y$ be finite-dimensional Hilbert
spaces. Assuming that we have universal sets of generators for the unitary groups $U(X)$
and $U(Y)$, the problem is to extend them to the universal set of generators for the
unitary group $U(X\otimes Y)$. First let us recall some useful facts from the
representation theory.

Let $V$ be a finite-dimensional Hilbert space and $\pi:G\To U(V)$ be a unitary
representation of a Lie group $G$. For $g\in G$ and $v\in V$ we shall use the notation
$gv$ for the action $\pi(g)(v)$. Let us denote by $C(\pi)$ the set of all such linear
operators $A:V\To V$ that $A\circ\pi(g)=\pi(g)\circ A,\ \forall\ g\in G$. It is clear
that $C(\pi)$ is a subalgebra of the algebra of all endomorphisms of $V$.
\begin{lem}
The representation $\pi$ is irreducible if and only if the algebra $C(\pi)$ does not
contain a projection operator different from 0 and 1.
\end{lem}
\begin{proof}
If the representation $\pi$ is reducible then $V=V_1\oplus V_2$ where $V_1$ and $V_2$ are
proper non-trivial invariant subspaces. It is clear that the projector on the subspace
$V_1$ (or $V_2$) commutes with every $\pi(g),\ g\in G$. Conversely, if $P:V\To V$ is a
non-trivial projector on a proper subspace which commutes with all $\pi(g)$, then we have
$$
\forall\ x\in\image{P},\ \forall\ g\in G:\ P(gx)=gP(x)=gx\gamomd gx\in\image{P}
$$
Hence, the subspace $\image{P}$ is invariant for every $\pi(g)$ and therefore the
representation $\pi$ is not irreducible.
\end{proof}
\begin{lem}
The representation $\pi$ is irreducible if the algebra $C(\pi)$ does not contain a
self-adjoint operator different from 0 and 1.
\end{lem}
\begin{proof}
If $\pi$ is not irreducible then according to the previous lemma the algebra $C(\pi)$
contains a projection different from 0 and 1, which is a self-adjoint operator.
Conversely, if $C(\pi)$ contains a self-adjoint operator $A$ different from 0 and 1,
consider the spectral decomposition of $A$: $A=\sum\lambda_iP_i$, where $P_i$-s are
orthogonal projectors. From $Ag=gA$ follows that if $x\in\image{P_i}$ then we have
$$
A(gx)=gA(x)=\lambda_i\cdot gx\gamomd gx\in\image{P_i}
$$
This implies that each subspace $\image{P_i}$ is invariant for the representation $\pi$.
\end{proof}
The last lemma implies the following
\begin{thm}
The representation $\pi$ is irreducible if and only if the algebra $C(\pi)$ consists of
only scalar operators (i.e., operators of the form $\lambda\cdot\mathbf{1},\
\lambda\in\Complex$).
\end{thm}
\begin{proof}
Let $A\in C(\pi)$ be a non-scalar operator. Then at least one of the following two
self-adjoint operators $B_1=A+A^*$ and $B_2=i(A-A^*)$ is also non-scalar. According to
the previous lemma this implies that the representation $\pi$ is not irreducible which
contradicts to the assumption of the theorem.
\end{proof}
For two representations $\pi_1:G_1\To U(V_1)$ and $\pi_2:G_2\To U(V_2)$, where $G_1$ and
$G_2$ are Lie groups and $V_1$ and $V_2$ are Hilbert spaces, let us denote by
$\pi_1\otimes\pi_2$ the representation of $G_1\times G_2$ in $V_1\otimes V_2$ defined as
$$
(\pi_1\otimes\pi_2)(g_1,g_2)(x,y)=g_1x\otimes g_2y,\ g_1\in G_1,\ g_2\in G_2,\ x\in V_1,\ y\in V_2%
$$
\begin{thm}
If the spaces $V_1$ and $V_2$ are finite-dimensional then
$C(\pi_1\otimes\pi_2)=C(\pi_1)\otimes C(\pi_2)$.
\end{thm}
\begin{proof}
Any operator $T:V_1\otimes V_2\To V_1\otimes V_2$ can be represented as
$T=\jami{}{p}A_p\otimes B_p$ where $A_p$ is an endomorphism of $V_1$ and $B_p$ is an
endomorphism of $V_2$. The RH side of this equality can be reduced to the form when the
operators $B_p$ are linearly independent. If $T\in C(\pi_1\otimes\pi_2)$ then we have the
following
$$
[T,g_1\otimes g_2]=0,\quad\forall\ g_1\in G_1\textrm{ and }\forall\ g_2\in G_2
$$
This implies that
$$
[T,g\otimes1]=0,\quad\forall\ g\in G_1
$$
Expanding $T$ we obtain
$$
\jami{}{p}[A_p,g]\otimes B_p=0,\quad\forall\ g\in G_1
$$
Since the operators $B_p$ are linearly independent, we have that each $[A_p,g]$ is 0.
Therefore, each $A_p\in C(\pi_1)$, which implies that $T\in End(V_1)\otimes C(\pi_2)$. In
same manner we obtain that $T\in C(\pi_1)\otimes End(V_2)$. Thus, we have that $T\in
C(\pi_1)\otimes End(V_2)\bigcap End(V_1)\otimes C(\pi_2)$ and therefore $T\in
C(\pi_1)\bigcap C(\pi_2)$.
\end{proof}
\begin{cor}\label{tensor_irreducible}
if the representations $\pi_1$ and $\pi_2$ are irreducible then the representation
$\pi_1\otimes\pi_2$ is also irreducible.
\end{cor}
Now consider the unitary group $U(V_1\otimes V_2)$ where $V_1$ and $V_2$ are
finite-dimensional Hilbert spaces and $\dim(V_1)=m$, $\dim(V_2)=n$. First of all notice
that its Lie algebra is
$$
\liealg{u}(V_1\otimes V_2)=\liealg{u}(V_1)\otimes i \liealg{u}(V_2)
$$
because (see \cite{Brylinski}), obviously the Lie algebra $\liealg{u}(V_1)\otimes i \liealg{u}(V_2)$ %
with the commutator defined as
$$
[a_1\otimes ib_1,a_2\otimes ib_2]=a_2a_1\otimes b_2b_1-a_1a_2\otimes b_1b_2
$$
is a subalgebra of $\liealg{u}(V_1\otimes V_2)$. Also we have that
$$
\dim(\liealg{u}(V_1\otimes V_2))=(mn)^2=\dim(V_1)^2\dim(V_2)^2=\dim(\liealg{u}(V_1)\otimes i\liealg{u}(V_2))%
$$
Consider the subgroup of $U(V_1\otimes V_2)$ generated by the elements of the form
$g_1\otimes g_2,\ g_1\in U(V_1),\ g_2\in U(V_2)$. Let us denote this subgroup by
$U(V_1)\otimes U(V_2)$. The Lie algebra of this subgroup is the subalgebra of elements of
the form
$$
u_1\otimes 1+1\otimes u_2,\quad u_1\in\liealg{u}(V_1)\ u_2\in\liealg{u}(V_2)
$$
Such element is known as the Kronecker sum of $u_1$ and $u_2$.
\begin{rem}
The Kronecker sum of $u_1$ and $u_2$ can be written as $u_1\otimes i(-i1)-i1\otimes iu_2$
in accordance with the equality $\liealg{u}(V_1\otimes V_2)=\liealg{u}(V_1)\otimes i\liealg{u}(V_2)$. %
\end{rem}
Consider the decomposition
$$
\begin{array}{c}
\liealg{u}(V_1\otimes
V_2)=(i\Real\cdot1\oplus\liealg{su}(V_1))\otimes(\Real\cdot1\oplus i\cdot\liealg{su}(V_2))=\\ %
\\
=(i\Real\otimes1)\oplus(\liealg{su}(V_1)\otimes1)\oplus(1\otimes\liealg{su}(V_2))\oplus(\liealg{su}(V_1)\otimes i\cdot\liealg{su}(V_2))=\\ %
\\
=(\liealg{u}(V_1)\otimes1+1\otimes\liealg{u}(V_2))\oplus(\liealg{su}(V_1)\otimes i\cdot\liealg{su}(V_2)) %
\end{array}
$$
The first summand is obviously the Lie algebra of $U(V_1)\otimes U(V_2)\equiv G$. %
Let us denote this Lie algebra by $L$. Consider the adjoint action of the group $G$ on
the Lie algebra $\liealg{u}(V_1\otimes V_2)$. It is clear that $L$ is invariant under
this action. Since the adjoint action of $U(V_1)$ on $\liealg{su}(V_1)$ is irreducible
and the same is true for $U(V_2)$ and $i\cdot\liealg{su}(V_2)$, we have according to the
corollary \ref{tensor_irreducible} that the action of $G$ on $\liealg{su}(V_1)\otimes
i\cdot\liealg{su}(V_2)$ is irreducible too. This together with Lemma
\ref{lemma_extension_of_generators} implies that if we have a set of generators of $L$ we
need only one new element from $\liealg{su}(V_1)\otimes i\cdot\liealg{su}V_2$ to generate
the entire Lie algebra $\liealg{u}(V_1\otimes V_2)$ (actually, any one element
\emph{outside} of $L$ is enough). Hence an optimal (in the sense of quantity) extension
of the set of generators of the Lie algebra of $U(V_1)\otimes U(V_2)$ to the set of
generators of the Lie algebra of $U(V_1\otimes V_2)$ is obtained by adding one and only
one element outside of the Lie algebra of $U(V_1)\otimes U(V_2)$.

The set of generators of Lie algebra is used to construct a control system the
corresponding set reachable from 1 of which is the entire Lie group. But the dynamics of
a control system is a continuous process while the conventional computational process
(quantum too) is discrete. In other words, the latter uses not 1-parameter flows of the
form $\exp(tX_i),\ t>0$ as gates to generate group elements, but discrete ''flows`` of
the form $g^n_i,\ n=1,2,\ldots$, where $g_i$-s are so called \emph{gates}. The generators
of Lie algebra can be used to construct such gates.

\newcommand{\gens}{\mathcal{X}}

Let $\gens=\set{X_1,\ldots,X_m}$ be a set of generators of the Lie algebra $L$ of a
compact connected Lie group $G$. For any $X_i\in\gens$ consider the corresponding
1-parameter flow $\exp(tX_i),\ t\in\Real$. If $\exp(tX_i)$ is periodic and therefore
$\exp(TX_i)=1$ for some $T\in\Real^+$ then take the element $g_i=\exp(\alpha_iTX_i)$,
where $\alpha_i\in\Real^+$ is an irrational number. The set $S(g_i)=\set{g_i^n\ |\
n\in\mathbb{N^+}}$ is dense in the orbit $\exp(tX_i),\ t\in\Real$. If for some generator
$X_j\in\gens$ the corresponding orbit $\exp(tX_j),\ t\in\Real^+$ is not periodic then
consider the maximal torus $\mathbb{T}_k\subset G$ where it is contained. replace the
element $X_j$ in the set of generators $\gens$ by periodic generators of the Lie algebra
of the torus $\mathbb{T}^k$. The obtained set is again a set of generators of $L$ because
$X_j$ is a linear combination of the generators of the torus. In such way we can achieve
that all the generators of $L$ are periodic, and then construct elements of
the group $G$: $g_1,\ldots,g_m$, such that for each of them the group $S(g_k)=\set{g_k^n\ |\ n\in\mathbb{N}^+},\ k=1,\ldots,m$ %
is dense in the 1-parameter subgroup $\exp(tX_k),\ t\in\Real$. Since the subgroups
$\exp(tX_k),\ k=1,\ldots,m$ generate the group $G$, the subgroups $S(g_k),\ k=1,\ldots,m$
generate a dense subgroup in $G$.
\begin{lem}
If $G$ is a Lie group with representation in some Hilbert space $\mathcal{H}$ and
$\mathcal{V}\subset\mathcal{H}$ is a subspace invariant for some dense subset $S\subset G$ %
then $\mathcal{V}$ is invariant for the entire group $G$.
\end{lem}
\begin{proof}
Consider the continuous map $F:G\To Hom(\mathcal{V},\mathcal{V}^\perp)$ defined as $F(g)=P\circ g|_{_\mathcal{V}}$ %
where $P:\mathcal{H}\To\mathcal{V}^\perp$ is the operator of orthogonal projection on
$\mathcal{V}^\perp$. Since $\mathcal{V}$ is invariant for $S$, we have that $F(S)=0$.
Since $S$ is dense in $G$ and $F$ is continuous, we have that $F(G)=0$ which means that
$\mathcal{V}$ is invariant for the entire group $G$.
\end{proof}
Let $L_1$ be a Lie subalgebra of $L$ and $G_1\subset G$ be the corresponding connected
Lie group. Let $L=L_1\oplus A$ be such decomposition of $L$ that the subspace $A$ is
invariant under the adjoint action of $G_1$ and the action of $G_1$ on $A$ is
irreducible. As it follows from Lemma \ref{lemma_extension_of_generators}, if we take
only one element $X\in A$, then the set $L_1\cap\set{X}$ generates the entire Lie algebra
$L$. If $X$ is such element that the 1-parameter subgroup $\set{\exp(tX)\ |\ t\in\Real}$
is periodic, then for some $\alpha\in\Real$, the element $g=\exp(\alpha X)$ generates a
dense subgroup in $\set{\exp(tX)\ |\ t\in\Real}$. In this situation the set
$G_1\cap\set{g}$ generates a dense subgroup in $G$. Moreover we the following
\begin{thm}
Under the above conditions the set $G_1\cap\set{g}$ generates the entire group $G$.
\end{thm}
\begin{proof}
First let us show that if $g_0=g^m$ is a normalizer of the subgroup $G_1$ then $m=0$.
Assume that $m>0$ and $g_0G_1g_0^{-1}=G_1$. It is clear that in this case the subgroup
$\set{g_0^n\ |\ n\in\mathbb{Z}}$ is also a normalizer of $G_1$. But this subgroup is
dense too in $\set{\exp(tX)\ |\ t\in\Real}\equiv S$. Consider the adjoint action of $S$
on the Lie algebra $L$. Since $\set{g_0^n\ |\ n\in\mathbb{Z}}$ is a normalizer of $G_1$,
we have that the group $S$ contains a dense subgroup for which $L_1$ is invariant. This
implies according to the previous lemma that $L_1$ is invariant for the adjoint action of
the group $S$. Therefore we have that $[X,L_1]\subset L_1$. Since $A$ is invariant under
the adjoint action of $G_1$ and $X\in A$, we have that $[X,L_1]\subset A$. The result is
that $[X,L_1]=0$, which contradicts to the assumption that the action of $G_1$ on $A$ is
irreducible. Hence, we obtain that there is no nontrivial normalizer for $G_1$ in
$\set{g^n\ |\ n\in\mathbb{Z}}$.

Now take any element $g_1=g^m,\ m\neq0$ and consider the group $G_2=g_1G_1g_1^{-1}$.
Since $g_1$ cannot be a normalizer of $G_1$, we have that $G_2\neq G_1$. Therefore, the
group $\hat{G}$ generated by $G_1\cap G_2$ contains $G_1$ as a proper subgroup. The group
$\hat{G}$ is connected because $G_1$ is such. Therefore, its Lie algebra $\hat{L}$
contains $L_1$ as a proper subalgebra. From this follows that $\hat{L}\cap A\neq0$. But
as the adjoint action of $G_1$ on $A$ is irreducible and $\hat{L}$ is invariant under
this action, we have that $\hat{L}\cap A=A$. The latter implies that $\hat{L}=L$. Since
$G$ and its subgroup $\hat{G}$ are connected and they have a common Lie algebra, they
must be equal: $\hat{G}=G$.
\end{proof}
Regarding the above theorem in the context of inductive construction of generators of the
unitary group of tensor product, we have the following situation: consider the unitary
group $U(V_1\otimes\cdots\otimes V_m)$, where $V_1,\ldots,V_m$ are some
finite-dimensional Hilbert spaces; if we are able to generate the subgroups
$U(V_1),\ldots,U(V_m)$, then we can construct a set of generators of the entire group
step-by-step by adding only one element at each step. For instance, consider the group
$U(V_1\otimes V_2\otimes V_3\otimes V_4\otimes V_5$. If we have the groups %
$U(V_1),\ U(V_2),\ U(V_3),\ U(V_4)\textrm{ and }U(V_5)$, then to construct %
$U(V_1\otimes V_2)$ and $U(V_3\otimes V_4)$ we need only 2 more elements. Then with
$U(V_1\otimes V_2)$ and $U(V_3\otimes V_4)$ we can construct %
$U(V_1\otimes V_2\otimes V_3\otimes V_4)$ by adding 1 more element. And finally, with
$U(V_1\otimes V_2\otimes V_3\otimes V_4)$ and $U(V_5)$ we can construct the entire group
$U(V_1\otimes V_2\otimes V_3\otimes V_4\otimes V_5)$ by adding again 1 more element.
Therefore, besides the subgroups $U(V_1),\ U(V_2),\ U(V_3),\ U(V_4)\textrm{ and }U(V_5)$
we need 4 more elements to construct the group $U(V_1\otimes V_2\otimes V_3\otimes V_4\otimes V_5)$.%

The above construction is, of course, in accordance with \cite{Brylinski}, and the key
point here is that an element outside of the subgroup is not a normalizer of this
subgroup.
\section{Control Systems on Homogeneous Spaces and Grassmann Manifold}
Let $M$ be a smooth manifold and $G$ be a Lie group acting on $M$. Let $\liealg{g}$ be
the Lie algebra of $G$. Any element $X\in\liealg{g}$ generates a vector field $\pi(X)$ on
$M$ defined as follows: for $m\in M$ let
$$
\pi(X)_m=R'_m(1)X
$$
where $R_m:G\To M$ is the map $R_m(g)=gm,\ g\in G$. We shall use the notation $Xm$ for
the value of the vector field $\pi(X)$ at a point $m\in M$. The vector field $\pi(X)$ is
not invariant under the action of $G$ and its conversion rule under this action is the
following: for $g\in G$ we have
$$
g(Xm)=gXg^{-1}gm,\quad\forall\ m\in M\gamomd g(\pi(X))=\pi(gXg^{-1})
$$
We consider control systems on $M$ of the form
\begin{equation}\label{homogeneous_contrsys}
\dot{m}=\pi(X_u)_m,\quad m\in M
\end{equation}
Where $X_u\in\liealg{g}$ and $u$ is a control parameter. Any solution of the system
\ref{homogeneous_contrsys} is contained in one orbit of the action of $G$ on $M$. This
fact easily follows from the definition of the vector field $\pi(X)$: for any point $m\in M$ %
the vector $\pi(X)_m$ is tangent to the orbit of the point $m$. Though the vector field
$\pi(X)$ is not invariant under the action of $G$ but we have the following
\begin{lem}
The integral manifolds of the differential system on $M$ generated by the mapping
$\liealg{g}\ni X\mapsto\pi(X)$ are exactly the orbits of the action of $G$.
\end{lem}
\begin{proof}
We apply the theorem of Nagano-Sussmann (see \ref{orbit_theorem_nagano-sussmann}) to this
case. For any $X\in\liealg{g}$ the action of the 1-parameter flow $\exp(tX),\ t\in\Real$
on the manifold $M$ generates the vector field which is exactly the same as the vector
field $\pi(X)$. For any $Y\in\liealg{g}$, by the conversion rule of the vector field
$\pi(Y)$ under the action of $G$ we obtain:
$$
\exp(tX)\pi(Y)=\pi(\exp(tX)Y\exp(-tX))
$$
which implies that $\pi([X,Y])=[\pi(X),\ \pi(Y)]$. Hence, the differential system
$m\mapsto\pi(\liealg{g})_m,\ m\in M$ is closed under the bracket operation. The
1-parameter flows of the vector fields $\pi(X),\ x\in\liealg{g}$ (as it was mentioned
just before) are $m\mapsto\exp(tX)m,\ m\in M$, and again according to the conversion rule
we have that for any $g\in G$:
$$
g\pi(\liealg{g})=\pi(g\liealg{g}g^{-1})=\pi(L)
$$
Therefore, the differential system $\pi(\liealg{g})$ is invariant under the action of
$G$, which implies according to the theorem of Nagano-Sussmann that the orbits of the
action of $G$ are the integral manifolds of the differential system $\pi(\liealg{g})$.
\end{proof}
From these discussions easily follows that a control system on homogeneous space has one
significant property: the complete controllability of such system is local (moreover: it
is pointwise): if $G$ acts on $M$ and the set of vector fields $\pi(V)$ generated by some
$V\subset\liealg{g}$ is completely nonholonomic in a point $m_0\in M$, then $\pi(V)$ is
completely nonholonomic in any point $m\in M$.

Now we consider the problem of controllability of finite-dimensional quantum systems in
the framework of control systems on homogeneous spaces. The classical notions of
\emph{transformation} and \emph{state} controllability can be carried to such systems. In
this case the corresponding Lie group is the unitary group $U(n)$ and the control system
$X_u$ is a parameter-dependent element of the Lie algebra $\liealg{u}(n)$.

The notion of \emph{state} in quantum (as well as in classical) case has various
meanings. The most general of them is \emph{density matrix}, i.e., a positive
self-adjoint matrix with trace equal to 1. The equation for the control system on the set
of density matrices is of the form
$$
\dot{\rho}=[X_u,\ \rho]
$$
where $\rho$ is a density matrix. Since any solution of such system belongs to one orbit
of the adjoint action of $U(n)$: $\rho(t)\in\set{g\rho_0g^{-1}\ |\ g\in U(n)}$, let us
recall the structure of an orbit of density matrix. For any density matrix $\rho$
consider its spectral decomposition
$$
\rho=\jami{i=1}{k}\lambda_iP_i,\quad\lambda_i>0,\ \jami{i=1}{k}\lambda_i=1,\ P_i^*=P_i,\ P_i^2=P_i,\ i=1,\ldots,k %
$$
The spectrum of the density matrix (operator) $\rho$ is the set of pairs
$$
Spec(\rho)=\set{(\lambda_i,\ m_i=rank(P_i))\ |\ i=1,\ldots,k}
$$
Two density matrices $\rho_1$ and $\rho_2$ are in one and the same orbit of adjoint
action if and only if $Spec(\rho_1)=Spec(\rho_2)$. Clearly, it is more reasonable to
consider the controllability not for the entire set of density matrices but for each
particular orbit of the adjoint action of $U(n)$.
\begin{defn}
A system $X_u\in\liealg{u}(n)$ is said to be \emph{state controllable} for a given orbit
$S$ of the adjoint action of $U(n)$ on the set of density matrices, if for any pair of
density matrices $\rho_0,\ \rho_1\in S$ there exists an admissible control $u$ and a time
$T\geq0$, such that the solution of $\dot{\rho}=[X_u,\ \rho],\ \rho(0)=\rho_0$, satisfies
$\rho(T)=\rho_1$.
\end{defn}
We shall formulate some criteria of state controllability for the special case when the
orbit of the adjoint action is the Grassmann manifold $Gr_k(\Complex^n)$. This is
equivalent to the case when $\lambda_i=1/k$ and $m_i=1$ $i=1,\ldots,k$. For any linear
operator $A:\Complex^n\To\Complex^n$ let $A_{12}$ be the upper-right component of the
decomposition
$$
A=
\begin{pmatrix}
  A_{11} & A_{12} \\
  A_{21} & A_{22} \\
\end{pmatrix}
$$ corresponding to the decomposition $\Complex^n=\Complex^k\oplus\Complex^{n-k}$. For
a given subset $\mathfrak{S}\subset\liealg{u}(n)$ let $Lie(\mathfrak{S})$ be the Lie
subalgebra of $\liealg{u}(n)$ generated by $\mathfrak{S}$.
\begin{thm}
Let $H:\Real^m\To\liealg{u}(n)$ be a continuous map and $C(\Real_+,\ \Real^m)$ be the
space of piecewise continuous functions from $\Real_+$ to $\Real^m$. For a given
Grassmann manifold
$$
Gr_k(\Complex^n)=\set{P:\Complex\To\Complex\ |\ P^*=P,\ P^2=P,\ trace(P)=k}
$$
the control system
$$
\dot{P}=[H(u),P],\quad P\in Gr_k(\Complex^n),\ u\in C (\Real_+,\ \Real^m)
$$
is controllable if and only if for the subspace $\Complex^k\To\Complex^{n-k}$ the set of
linear maps
$$
X_{12}:\Complex^k\To\Complex^{n-k},\quad X\in Lie(\image{H})
$$
gives the entire space $Hom(\Complex^k,\ \Complex^{n-k})$.
\end{thm}
The proof follows from the already mentioned fact that the necessary and sufficient
condition for the complete controllability of a control system on a homogeneous space is
that the corresponding differential system be completely nonholonomic in at least one
point, and the fact that the tangent space of the Grassmann manifold $Gr_k(\Complex^n)$
at a point $X$ is canonically isomorphic to $Hom(X,\ X^\perp)$.
\bibliographystyle{amsplain}

\end{document}